\documentclass[%
twocolumn,
 reprint,
superscriptaddress,
 amsmath,amssymb,
 aps,
 rc ]{revtex4-2}
\usepackage{comment}

\usepackage{graphicx}
\usepackage{dcolumn}
\usepackage{bm}
\usepackage{xcolor}
\usepackage{siunitx}

\usepackage{hyperref}

\begin{document}

\preprint{APS/123-QED}

\title{Indistinguishable photons from an artificial atom in silicon photonics}

\author{Lukasz Komza}
\thanks{L.K. and P.S. contributed equally.}
\affiliation{
Department of Physics, University of California, Berkeley, Berkeley, California 94720, USA
}
\affiliation{
 Materials Sciences Division, Lawrence Berkeley National Laboratory, Berkeley, California 94720, USA
}%
\author{Polnop Samutpraphoot}
\thanks{L.K. and P.S. contributed equally.}
\affiliation{
 Materials Sciences Division, Lawrence Berkeley National Laboratory, Berkeley, California 94720, USA
}%
\affiliation{
Department of Electrical Engineering and Computer Sciences, University of California,  Berkeley, Berkeley, California 94720, USA
}
\author{Mutasem Odeh}
\affiliation{
 Materials Sciences Division, Lawrence Berkeley National Laboratory, Berkeley, California 94720, USA
}%
\affiliation{
Department of Electrical Engineering and Computer Sciences, University of California,  Berkeley, Berkeley, California 94720, USA
}
\author{Yu-Lung Tang}
\affiliation{
Department of Physics, University of California, Berkeley, Berkeley, California 94720, USA
}
\affiliation{
 Materials Sciences Division, Lawrence Berkeley National Laboratory, Berkeley, California 94720, USA
}%
\author{Milena Mathew}
\affiliation{
 Materials Sciences Division, Lawrence Berkeley National Laboratory, Berkeley, California 94720, USA
}%
\affiliation{
Department of Electrical Engineering and Computer Sciences, University of California,  Berkeley, Berkeley, California 94720, USA
}
\author{Jiu Chang}
\affiliation{
Department of Electrical Engineering and Computer Sciences, University of California,  Berkeley, Berkeley, California 94720, USA
}
\author{Hanbin Song}
\affiliation{
Department of Materials Science and Engineering, University of California, Berkeley, Berkeley, California 94720, USA
}
\author{Myung-Ki Kim}
\affiliation{
Department of Electrical Engineering and Computer Sciences, University of California,  Berkeley, Berkeley, California 94720, USA
}
\author{Yihuang Xiong}%

\affiliation{%
 Thayer School of Engineering, Dartmouth College, 14 Engineering Dr, Hanover, NH 03755, USA 
}%

\author{Geoffroy Hautier}
\affiliation{%
 Thayer School of Engineering, Dartmouth College, 14 Engineering Dr, Hanover, NH 03755, USA
}%

\author{Alp Sipahigil}
\email{Corresponding author: alp@berkeley.edu}

\affiliation{
Department of Electrical Engineering and Computer Sciences, University of California,  Berkeley, Berkeley, California 94720, USA
}
\affiliation{
 Materials Sciences Division, Lawrence Berkeley National Laboratory, Berkeley, California 94720, USA
}%
\affiliation{
Department of Physics, University of California, Berkeley, Berkeley, California 94720, USA
}

%




\date{\today}

\begin{abstract}

%

Silicon is the ideal material for building electronic and photonic circuits at scale. Spin qubits and integrated photonic quantum technologies in silicon offer a promising path to scaling by leveraging advanced semiconductor manufacturing and integration capabilities. However, the lack of deterministic quantum light sources, two-photon gates, and spin-photon interfaces in silicon poses a major challenge to scalability. In this work, we show a new type of indistinguishable photon source in silicon photonics based on an artificial atom. We show that a G center in a silicon waveguide can generate high-purity telecom-band single photons. We perform high-resolution spectroscopy and time-delayed two-photon interference to demonstrate the indistinguishability of single photons emitted from a G center in a silicon waveguide. Our results show that artificial atoms in silicon photonics can source highly coherent single photons suitable for photonic quantum networks and processors.  

\end{abstract}

\maketitle


Silicon quantum technologies based on spin qubits~\cite{Zwanenburg2013} and integrated photonics~\cite{Wang2019} offer a promising path to scaling by leveraging advanced semiconductor manufacturing and integration capabilities~\cite{Zwerver2022,Sun2015microprocessor}. 
Current approaches to fault-tolerant  photonic quantum computation use weak material nonlinearities and measurements to probabilistically generate photon pairs and implement two-qubit gates~\cite{Silverstone2013,Wang2019}. The lack of deterministic quantum light sources \cite{Kuhn2002singlephotonsource}, photon-photon gates \cite{DuanKimble2004photonphoton, Hacker2016photonphoton}
, and quantum memories \cite{Specht2011memories} in silicon photonics poses a major challenge to scalability and results in very large resource overheads \cite{Li2015}. Coherently controlled quantum emitters in a reconfigurable photonic circuit can enable hardware-efficient universal quantum computation~\cite{Pichler2017,Bartlett2021} and time-multiplexed quantum networking \cite{Pu2017}. Silicon photonics provides a mature platform for low-loss reconfigurable integrated photonics~\cite{Zhang2022}. However, an atomic source of indistinguishable photons in silicon has been missing~\cite{Yan2021}. We address this challenge by demonstrating telecom-band indistinguishable photon generation from an artificial atom in silicon photonics.

\begin{figure}[t!]
 	\centering
 	\includegraphics[width=0.5\textwidth]{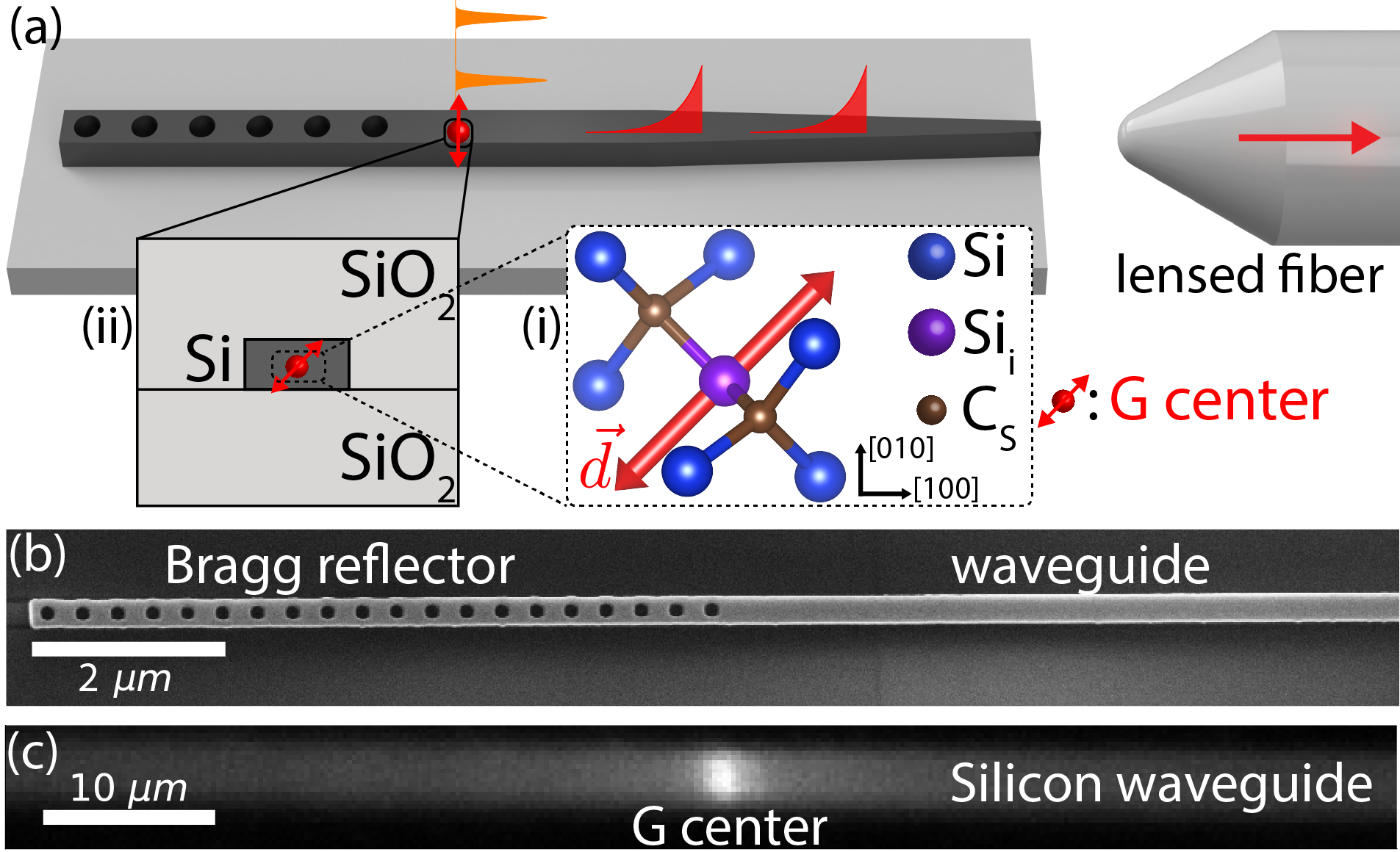}
 	\caption{\textbf{An atomic quantum light source in silicon photonics. } 
  (a) Device and measurement schematic. (i) A G center is created inside a silicon photonic waveguide via ion implantation. (ii) Silicon waveguide cross section. G center transition dipole $\vec{d}$ is along $\langle110\rangle$. Laser excitation (orange) of the G center results in single-photon emission into the waveguide (red) which is collected using a lensed fiber. (b) Scanning electron micrograph of the photonic waveguide near the broadband Bragg reflector. (c) PL image of the waveguide shows bright emission from an isolated G center. }
 	\label{fig1}
\end{figure}

Artificial atoms in solids enable single-photon level optical nonlinearities for realizing deterministic single-photon sources, two-photon gates, and long-range spin-spin entanglement~\cite{Atatre2018,Sipahigil2016,bhaskar2020mdi, Knall2022efficient, Bernien2013entanglement, Pompili2021}. 
While defect-based photoluminescence (PL) in silicon has been studied for decades \cite{Davies1989}, bright telecom-band single-photon emission from a broad diversity of artificial atoms in silicon was only recently shown \cite{Redjem2020telecom,Durand2021, prabhu2022addressable, Higginbottom2022}. In order for  silicon artificial atoms to function as quantum-coherent light sources, their emission has to satisfy  spatiotemporal indistinguishability~\cite{Legero2003}. In this work, we integrate a silicon color center into a photonic waveguide, show pulsed single-photon generation, and demonstrate that successive photons emitted are indistinguishable.

\noindent{}\textbf{An artificial atom in a silicon waveguide.} Our device consists of a  G center created in a silicon photonic waveguide (Fig.~\ref{fig1}). 
The G center is a complex defect in silicon that consists of two substitutional carbon ($\textrm{C}_\textrm{S}$) atoms and an interstitial silicon ($\textrm{S}_\textrm{i}$). It emits in the telecommunication O-band with a zero-phonon line (ZPL) at 1278~nm \cite{Cloutier2005, Murata2011}. 
We create G centers inside the \SI{220}{nm} device layer of a silicon-on-insulator (SOI) wafer by ${}^{12}\textrm{C}$ ion implantation at 36 keV and a fluence of $10^{12}\,\mathrm{cm}^{-2}$, followed by rapid thermal annealing at 1000°C for 20 seconds. These parameters result in the creation of approximately one G center in a  $\SI{100}{\um}$-long waveguide. 

\begin{figure}
    \centering
    \includegraphics[width=\columnwidth]{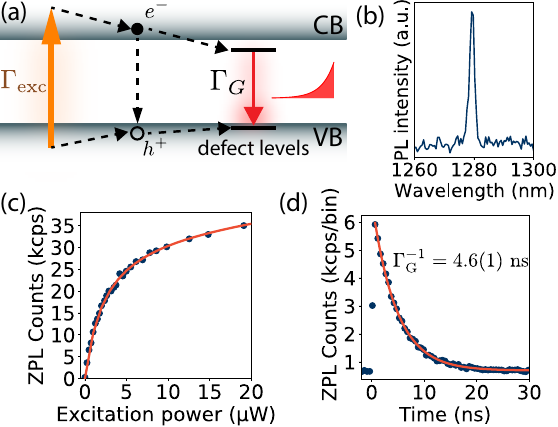}
    \caption{
        \textbf{Photodynamics of a G center in a waveguide.} 
          (a) Above-bandgap excitation ($\Gamma_\textrm{exc}$) creates excess carriers that recombine at the localized defect levels ($\Gamma_\textrm{G}$), and produce single-photon emission. (b) PL spectrum of the ZPL of the G center. 
          (c) Excitation power dependence of the ZPL PL shows saturated emission.  
          (d) PL lifetime measurement with pulsed excitation. 
  }
 	\label{fig2}
\end{figure}

Upon above-bandgap excitation, the G center emits photons via the radiative recombination of electron-hole pairs at localized defect levels (Fig.~\ref{fig2}(a)). The dipole emission is guided by a single-mode silicon waveguide coupled to a single-mode lensed fiber with $ 50\%$ efficiency using an adiabatic mode converter (Fig.~\ref{fig1}(a),\cite{SI}). The waveguide is terminated with a Bragg reflector for single-sided measurements and coupling efficiency calibration. 
The collected photons are detected using a spectrometer (Fig.~\ref{fig2}(b)) or superconducting nanowire single-photon detectors (SNSPD) with a quantum efficiency of $60 \%$. The sample is housed in a cryostat and measured at \SI{3.4}{K}. Materials, fabrication, photonic design, setup, and first principles calculation details are provided in \cite{SI}.

\noindent{}\textbf{Optical properties of a G center in a waveguide. }
 To locate a single G center, 
we spatially scan a free-space excitation beam at 635~nm and detect photons emitted into the waveguide through the lensed fiber.  
Fig.~\ref{fig1}(c) shows the resulting PL image of the waveguide where we observe an isolated emitter with a measured photon rate of $18$~kcps using a bandpass filter ($1280\pm6$~nm) 
centered at the G center ZPL (emission spectrum shown in Fig.~\ref{fig2}(b)). In the following experiments, we probe the linear and nonlinear optical responses of this G center using time- and spectrally- resolved single-photon detection.

\begin{figure}[t!]
 	\centering
 	\includegraphics[width=\columnwidth]{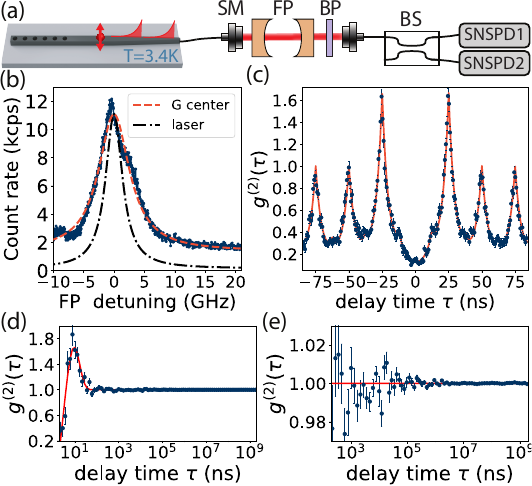}    
 	\caption{\textbf{Quantum coherence of G center emission.} 
  (a) Emission from the G center is collected through a single mode (SM) fiber and analyzed with a tunable Fabry Perot (FP) cavity. BP: Bandpass filter, BS: beamsplitter.
  (b) Measured emission linewidth (red fit): \SI{6.2\pm 0.1}{GHz}, FP linewidth (black, using reference laser): \SI{3.4\pm 0.1}{GHz}. Calculated G center linewidth after deconvolving the FP response: \SI{2.8\pm 0.1}{GHz}.
 (c) Normalized intensity correlations at the detectors $g^{(2)}(\tau)$ show single-photon emission $g^{(2)}(0) = 0.15(2) <0.5$. (d,e) Long term intensity correlations under continuous wave excitation show stable single-photon emission.
}
 	\label{fig3}
\end{figure}

\begin{figure*}[t!]
 	\centering
 	\includegraphics[width=2\columnwidth]{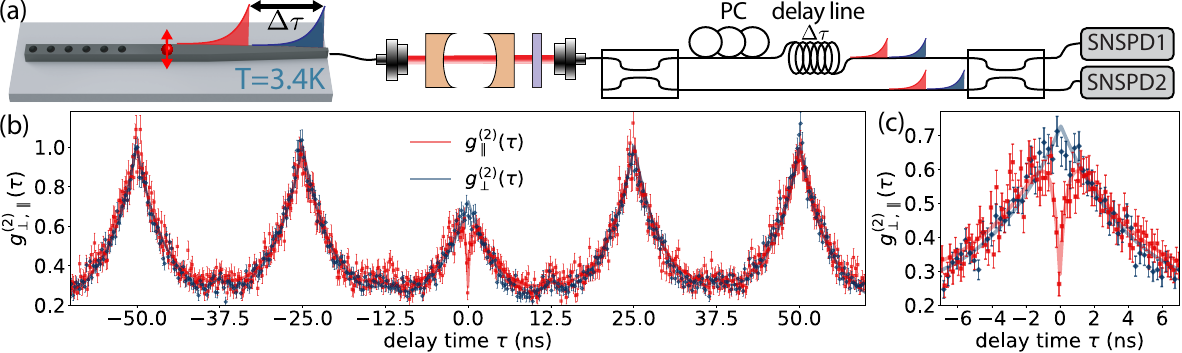}
 	\caption{\textbf{Quantum interference of single photons from a G center in silicon photonics. } 
  (a) Successive photons ($\Delta \tau = 25$~ns)  interfere via a time-delayed Mach-Zehnder interferometer. Indistinguishability of the interfering single photons is adjusted by setting their relative polarizations using a polarization controller (PC).
  (b) Two-photon interference measurement. 
  $g^{(2)}(\tau)$ for orthogonal (blue) and parallel (red)  polarizations. (c) When the two photons are indistinguishable, quantum interference results in the antibunching dip at short time delays. We observe nearly identical correlations outside of the quantum interference window.}
 	\label{fig4}
\end{figure*}

We study the saturation response of the G center by measuring the power dependence of the ZPL emission rate $R_{\textrm{ZPL}}$ on the excitation power $P$. The power dependence is modeled by $R_{\textrm{ZPL}}=R_\textrm{sat}/(1+P/P_\textrm{sat})+\alpha P, $
where the two terms correspond to a two-level atomic response and a weak linear background. The fit yields a saturated count rate of $R_\textrm{sat} = 35~$kcps and a saturation power $P_\textrm{sat} = \SI{2.4}{\uW}$ (Fig.~\ref{fig2}(c)).  Next, we use a pulsed laser at 705~nm to measure the PL lifetime of the emitter to be $\Gamma_G^{-1}= $\SI{4.6\pm0.1}{ns} (Fig.~\ref{fig2}(d)). 
For each excitation pulse, we detect a ZPL photon with a probability of  $0.4(1)\times 10^{-3}$  \cite{SI}. We calibrate the losses in our setup and use the ZPL branching ratio of 0.18 to estimate the probability the G center emits into the waveguide  $\beta=\Gamma_{1D}/\Gamma_G=0.014$, where $\Gamma_{1D}$ is the radiative emission rate into the waveguide. 
We estimate a radiative lifetime upper bound of \SI{260}{ns}. Our first principles calculations predict a radiative lifetime of \SI{225\pm75}{ns}~\cite{SI}. 

\noindent{}\textbf{Optical coherence.} We probe the optical coherence of the G center by measuring the ZPL emission spectrum using a tunable Fabry-Perot (FP) cavity with a linewidth of $\kappa_\textrm{FP}/2\pi = $ 3.4~GHz (Fig.~\ref{fig3}(a)). The resulting spectrum, which is a convolution of the ZPL emission and the FP transmission, shows a total linewidth of 6.2~GHz. After deconvolving the cavity response, we find the G center emission linewidth to be $\Gamma / 2\pi =$ 2.8~GHz (Fig.~\ref{fig3}(b)). 

Next, we characterize the photon statistics of the G center emission by measuring the normalized intensity correlations $g^{(2)}(\tau)$ under pulsed 705~nm excitation at a repetition period $\Delta \tau=25$~ns (Fig.~\ref{fig2}(c)). We observe antibunched intensity correlations at zero delay $g^{(2)}(0) = 0.15(2) <  0.5$ that confirm single-photon emission.  The  $g^{(2)}(0)$ value is limited by contributions from the  ratio of the repetition period and the excited state lifetime ($\Delta \tau /\Gamma_G^{-1}$), imperfect extinction in pulsed laser downsampling, and dark counts~\cite{SI}. We benchmark the long-term stability of the G center emission by analyzing the intensity correlations up to seconds of delays under CW excitation. The results in Fig.~\ref{fig3}(d,e) show a flat response which indicates stable single-photon emission without any excess intensity fluctuations for $\tau>\SI{50}{ns}$. We observe bunching at shorter timescales which has been attributed to the presence of a metastable state~\cite{Redjem2020telecom,prabhu2022addressable}. 

\noindent {\bf Time-resolved two-photon quantum interference}. Photon indistinguishability requires a high degree of spatio-temporal overlap between single-photon wavepackets emitted from the sources~\cite{Legero2003}. 
We use a time-delayed Hong-Ou-Mandel (HOM) interference experiment to test the indistinguishability of successive single-photon pulses from the G center \cite{Santori2002, Gazzano2013}. We interfere successive single photons (red and blue pulses in Fig.~\ref{fig4}(a)) using a fiber-based time-delayed Mach-Zehnder interferometer (MZI) where one path has an additional delay $\Delta \tau=25$~ns, matched to the laser repetition period. We adjust the relative polarization between the two MZI paths to control the mode overlap and photonic indistinguishability at the second beam splitter~\cite{SI}.

The results of HOM interference between parallel and orthogonally polarized single-photon pairs from a G center are shown in Fig.~\ref{fig4}(b,c). When the polarizations of the two interfering photons are parallel (red data, indistinguishable case), we see the characteristic HOM dip \cite{Hong1987} resulting from two-photon quantum interference at short time delays with $g^{(2)}_\parallel(0) = 0.26(4) < 0.5$. When we tune the photons to be orthogonally polarized so that they are intentionally distinguishable (blue data), the HOM dip disappears and we obtain $g^{(2)}_\perp(0) = 0.69(5)$. A comparison of the normalized coincidence probability at zero time delay yields an HOM interference visibility of $\chi=1-g^{(2)}_\parallel(0)/g^{(2)}_\perp(0)=0.62(4)$. The visibility is primarily limited by the timing jitter of our detector pair ($\SI{250}{ps}$) with minor contributions from imperfect polarization overlap, finite lifetime-to-repetition period ratio, and dark counts. The decay time of quantum interference, $\tau_\textrm{HOM} = 0.42(10)$~ns, is an order of magnitude shorter than the excited state lifetime $\Gamma_G^{-1}=\SI{4.6}{ns}$ but an order of magnitude longer from an estimate based on the measured G center linewidth $(2\Gamma)^{-1}= \SI{0.03}{ns}$ ~\cite{SI}. These results show that noise sources leading to optical decoherence have long-time correlations. The quantum interference decay time $\tau_\textrm{HOM}$ indicates an effective optical linewidth of $380$~MHz at short timescales.

Our results show that color centers in silicon can generate indistinguishable photons at the telecom-band in silicon photonics. This experiment is enabled by the large transition dipole moment ($2.8\pm0.5$ Debye calculated from first principles in \cite{SI}), the optical coherence of the G center, and efficient collection of single photons from the G center using silicon photonics. In the following, we discuss open questions and approaches to advance this platform to develop deterministic spin-photon interfaces for silicon-based quantum repeaters and integrated photonic quantum processors.

High-fidelity atom-photon and photon-photon gates necessary for quantum repeaters and processors require operation of artificial atoms in the high-cooperativity regime of cavity or waveguide quantum electrodynamics \cite{DuanKimble2004photonphoton,Sipahigil2016,bhaskar2020mdi}. In a one dimensional system, the cooperativity  is  given by $C=\Gamma_{1D}^{ZPL}/\Gamma'$ where $\Gamma_{1D}^{ZPL}$ is the ZPL emission rate into the waveguide and $\Gamma'$ is the sum of all other broadening mechanisms including emission into the phonon sideband, non-radiative decay, free space emission, and  optical decoherence due to spectral diffusion. Of these parameters, we find that spectral diffusion dominates the effective linewidth with $\Gamma'\approx(\tau_\textrm{HOM})^{-1}=2\pi \times~380$~MHz, similar to other solid-state emitters \cite{Atatre2018,Acosta2012}. 
We estimate  $\Gamma_{1D}^{ZPL}/2\pi\approx0.1-1 $ MHz where the large uncertainty is due to the random positioning of the G center along the single-sided waveguide. These estimates correspond to $C\sim 10^{-3}$ in the current experiment. 

The strategies to realize high-fidelity silicon quantum repeaters and processors with $C\gg1$ focus on enhancing $\Gamma_{1D}^{ZPL}$ and reducing $\Gamma'$. High $Q/V$ silicon  photonic resonators  \cite{Asano2017,Qubaisi2021} with mode volume $V\sim0.1\lambda^3$ and quality factor $Q\sim10^6$ can Purcell-enhance the coherent atom-photon interaction rate by $\sim10^6$. The spectral diffusion of the emitters can be suppressed to achieve lifetime-limited optical linewidths by using resonant excitation and dynamic stabilization \cite{Acosta2012}, as well as embedding emitters in p-n junctions where charge noise can be strongly suppressed \cite{Anderson2019}. Finally, the discovery of new centrosymmetric artificial atoms \cite{Durand2021} in silicon will make these systems more robust against spectral diffusion \cite{Sipahigil2014}. With such realistic improvements, a cooperativity of $C\sim10^4$ can be achieved to realize high fidelity atom-photon and photon-photon interactions above quantum error correction thresholds in silicon quantum photonics. Emerging silicon artificial atoms with electron and nuclear spins \cite{Higginbottom2022} can also be introduced to this  device platform to implement quantum processor and repeater building blocks, as was shown in other material platforms \cite{bhaskar2020mdi, Pompili2021, kalb2017distillation, Abobeih2022}. 
The realization of such high-fidelity spin-photon gates with silicon artificial atoms will open up the possibility of scaling spin~\cite{Zwanenburg2013} and photonic~\cite{Wang2019} quantum processors and repeaters using advanced CMOS manufacturing and integration capabilities~\cite{Zwerver2022,Sun2015microprocessor,Yan2021}.

\noindent{}\textbf{Acknowledgments.} 
This research was led by funding through NSF QuIC-TAQS program through award No 2137645. Additional support was provided by the U.S. Department of Energy, Office of Science, Basic Energy Sciences 
in Quantum Information Science under Award Number DE-SC0022289 for materials processing and first principles modeling, and the NSF Challenge Institute for Quantum Computation (CIQC). This research used resources of the National Energy Research Scientific Computing Center, a DOE Office of Science User Facility supported by the Office of Science of the U.S. Department of Energy under Contract No. DE-AC02-05CH11231 using NERSC award BES-ERCAP0020966. We thank UC Berkeley, LBNL, Ming Wu and Eli Yablonovitch for their support during the ramp up of our new lab, Quantum Opus for custom SNSPD installation, Auden Young, Andrew Kim, and Xudong Li for technical assistance, and Zihuai Zhang for feedback on the manuscript. The devices used in this work were fabricated at UC Berkeley's NanoLab.


\appendix
\renewcommand{\thefigure}{S\arabic{figure}}
\setcounter{figure}{0}

\section{Experimental setup}
\begin{figure*}[t!]
 	\centering
 	\includegraphics[width=1.6\columnwidth]{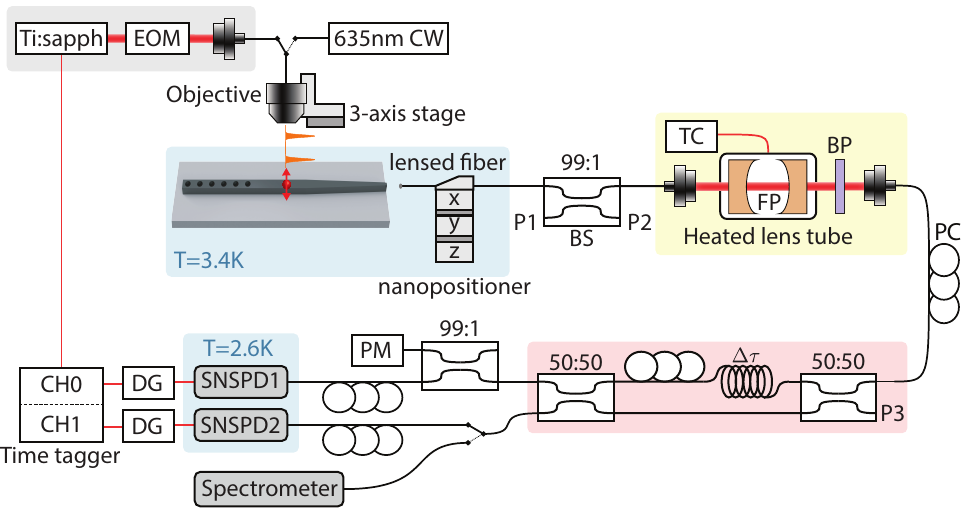}
 	\caption{\textbf{Experimental setup details.} Black and orange lines represent optical fiber and electrical connections respectively. Empty ports of fiber beamsplitters (P1, P2, P3) were used for efficiency calibrations, such as measuring the fiber coupling or transmission through the fabry-perot cavity. EOM: electro-optic modulator. 635~nm CW: 635~nm continuous-wave laser. BS: beamsplitter. TC: temperature controller. FP: Fabry Perot cavity. BP: bandpass filter. PC: polarization controller. PM: polarimeter. SNSPD: superconducting nanowire single-photon detector. DG: delay generator. CH0/CH1: time tagger channels. $\Delta\tau$ = 25~ns delay line.}
 	\label{fig:network}
\end{figure*}
The sample is mounted in a cryostat (Montana Instruments Cryostation s200) and cooled down to 3.4~K.  A lensed fiber (OZ Optics TSMJ-X-1550-9/125-0.25-7-2.5-14-2) used for photon collection through the waveguide is mounted on a 3-axis nanopositioner (Attocube ANPx101/LT and ANPz102/LT) used for fiber alignment. Photons are detected by a pair of SNSPDs (Quantum Opus QO-NPD-1200-1600), each with 60\% detection efficiency at optimal polarization. 

We calibrate our fiber coupling efficiency by injecting laser light into port P2 (Fig.~\ref{fig:network}) and measuring the reflected power after the first beamsplitter. We calculate the waveguide-fiber coupling efficiency $\eta_\mathrm{fc}$ from
$$P_\mathrm{out} = \left(\eta_\mathrm{fc}^2 RT\right) P_\mathrm{in}$$
where $R$ and $T$ are the measured reflection and transmission coefficiencts of the beamsplitter. We measure $\eta_\mathrm{fc} \approx 0.5$ for TE-polarized light. The photonic crystal end mirror is reflective for TE polarization, and maximizing the reflected power allows us to selectively excite TE-polarized light. 

The excitation laser beam is sent through the top vacuum window of the cryostat and focused by a microscope objective (Mitutoyo LCD Plan Apo NIR 50, NA=0.42) mounted on a 3-axis translation stage (Sutter Instrument MP-285) used for raster scanning. Measurements in Fig.~\ref{fig1}(c), Fig.~\ref{fig2}(b,c), and Fig.~\ref{fig3}(b,d,e) are performed with continuous wave 635~nm excitation (Thorlabs S1FC635), whereas measurements in Fig.~\ref{fig2}(d), Fig.~\ref{fig3}(c), and Fig.~\ref{fig4}(b) are performed with 705-715~nm pulsed Ti:sapphire laser (Coherent Chameleon Ultra II). The pulsed laser has a pulse duration of \SI{140}{fs} and repetition rate of 80 MHz. To perform experiments at \SI{40}{MHz}, the laser was downsampled using an electro-optic modulator to suppress every other pulse. We observed finite pulse suppression, with 8.0 dB extinction of suppressed pulses, resulting in weak but observable contributions in our measurements at odd multiples of the repetition period (Fig.~\ref{fig3}(c), Fig.~\ref{fig4}(b)).

We tuned the photon indistinguishability in the HOM experiment by aligning the relative polarization of the two arms of the MZI interferometer to be parallel and perpendicular. To achieve this, we artificially broadened a tunable O-band laser (Santec TSL-570) to eliminate interference effects while aligning the polarization. When we inject a broadband light source into the MZI interferometer in Fig.~\ref{fig4}, if the two arms have orthogonal polarizations, the output orthogonal polarization is completely unpolarized. If the polarizations are parallel the output state is completely polarized. Therefore, we measure the degree of polarization (DOP) and tune it to be either 0\% or 100\% using a fiber polarization controller in one arm of the interferometer. We use a polarimeter (ThorLabs PAX1000IR2) to measure the DOP of MZI output with broadband light. We achieved DOPs within 5\% of 0\% and 100\%, where the DOP drifted slowly due to polarization drifts from thermal fluctuations.

\section{Device fabrication and design}
\label{SI:fab}

\begin{figure}
    \centering
    \includegraphics[width=\columnwidth]{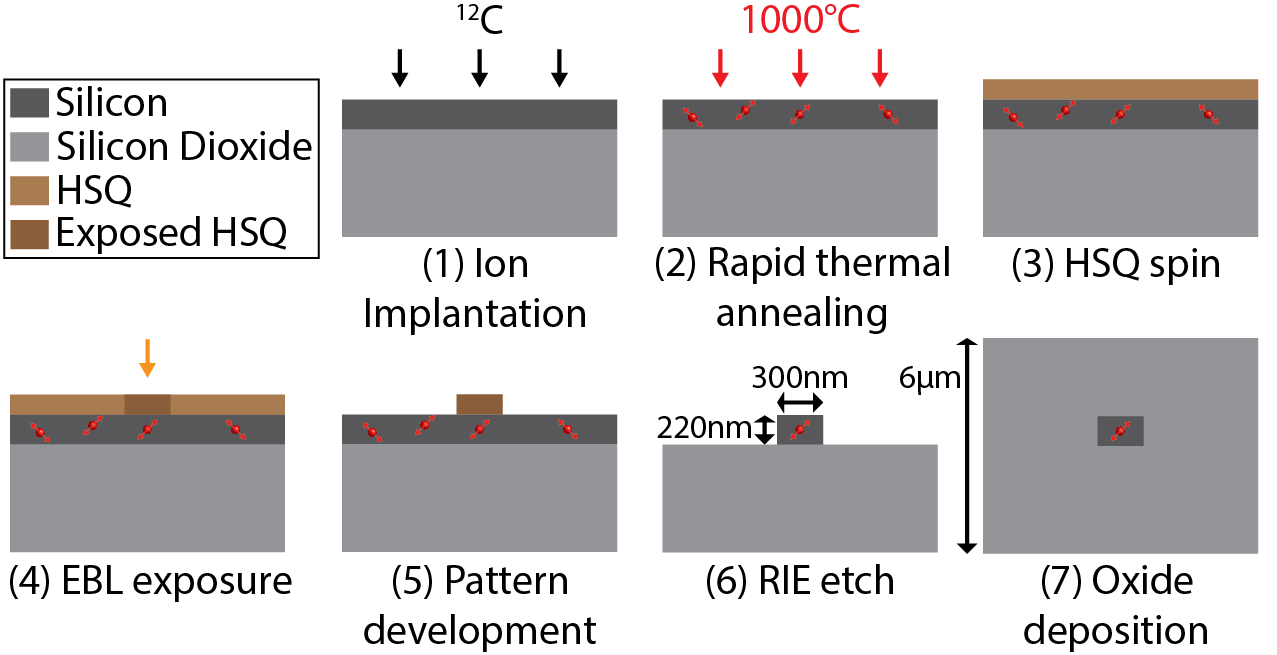}
    \caption{\textbf{Device fabrication process}. All photonics fabrication is done post-implantation. After oxide deposition, the waveguide facets are defined by cleaving the chip.}
    \label{figS1:fab}
\end{figure}

\begin{figure}
    \centering
    \includegraphics[width=\columnwidth]{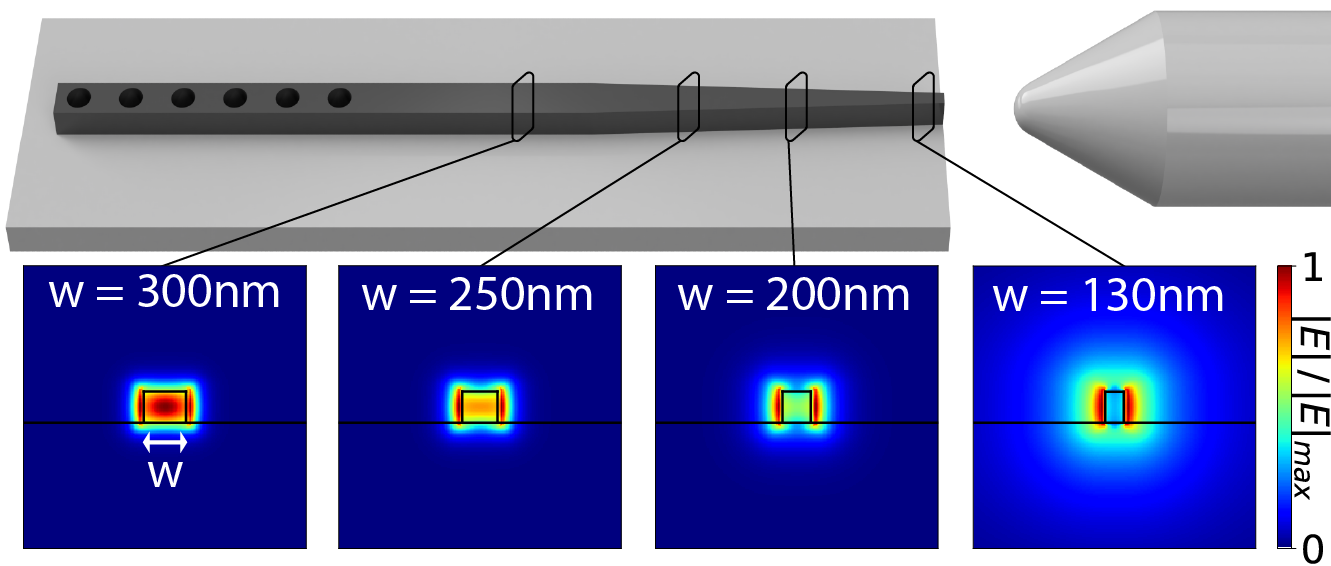}
    \caption{Adiabatic mode coupler. The waveguide width is tapered to mode match with the lensed fiber, which has a mode field diameter (MFD) of \SI{2.5}{\mu m}. The adiabatic tapering region is \SI{50}{\mu m} long to have high efficiency. The emitter shown in the manuscript is in the \SI{300}{nm} wide region to achieve stronger coupling to the waveguide. }
    \label{fig:taper}
\end{figure}

The sample was prepared from a 1~cm $\times$ 1~cm chip diced from a \SI{200}{mm}, high-resistivity (Float-zone, $\geq$\SI{3000}{\ohm cm}), \SI{220}{nm} SOI wafer prepared using the SmartCut method. The fabrication process is summarized in Fig.~\ref{figS1:fab}. 
The backside of the chip was partially diced to enable cleaving of the chip at desired locations to expose the waveguide facet for fiber coupling. Carbon implantation and high temperature annealing were carried out before any lithography steps. An etch mask for photonic structures was defined through electron beam lithography on HSQ resist. The mask was developed using a NaOH/NaCl developer chemistry. The 220~nm device layer was etched in a $\mathrm{Cl_2 / HBr}$ chemistry. Finally, approximately $\SI{3}{\um}$ of silicon oxide cladding was deposited on the surface of the chip through PECVD. The resulting waveguides have the cross section shown in Fig.~\ref{fig:taper}.

We used Finite Element Method (FEM) and Finite-Difference Time-Domain (FDTD) solvers to design a Bragg reflector for the fundamental TE mode based on a photonic crystal with lattice constant of 370~nm and ellipsoid holes with the principle axes of \{170, 200\}~nm. This design yields near-unity reflection for TE polarization over a 150~nm band centered at 1330~nm. The center of the band was biased towards wavelengths longer than the G center ZPL to include the phonon sideband. The waveguide width of 300~nm was optimized to achieve the maximum electric field intensity at the center of the waveguide, and therefore the coupling strength to the emitter.  The waveguide width was tapered down to 130~nm over $\SI{50}{\um}$ for maximum coupling to a $\SI{2.5}{\um}$ lensed fiber (Fig.~\ref{fig:taper}). The waveguide is fabricated along the $\langle 100 \rangle$ crystal axis.

\section{Carbon diffusion in silicon}

\begin{figure}
    \centering
    \includegraphics[width=\columnwidth]{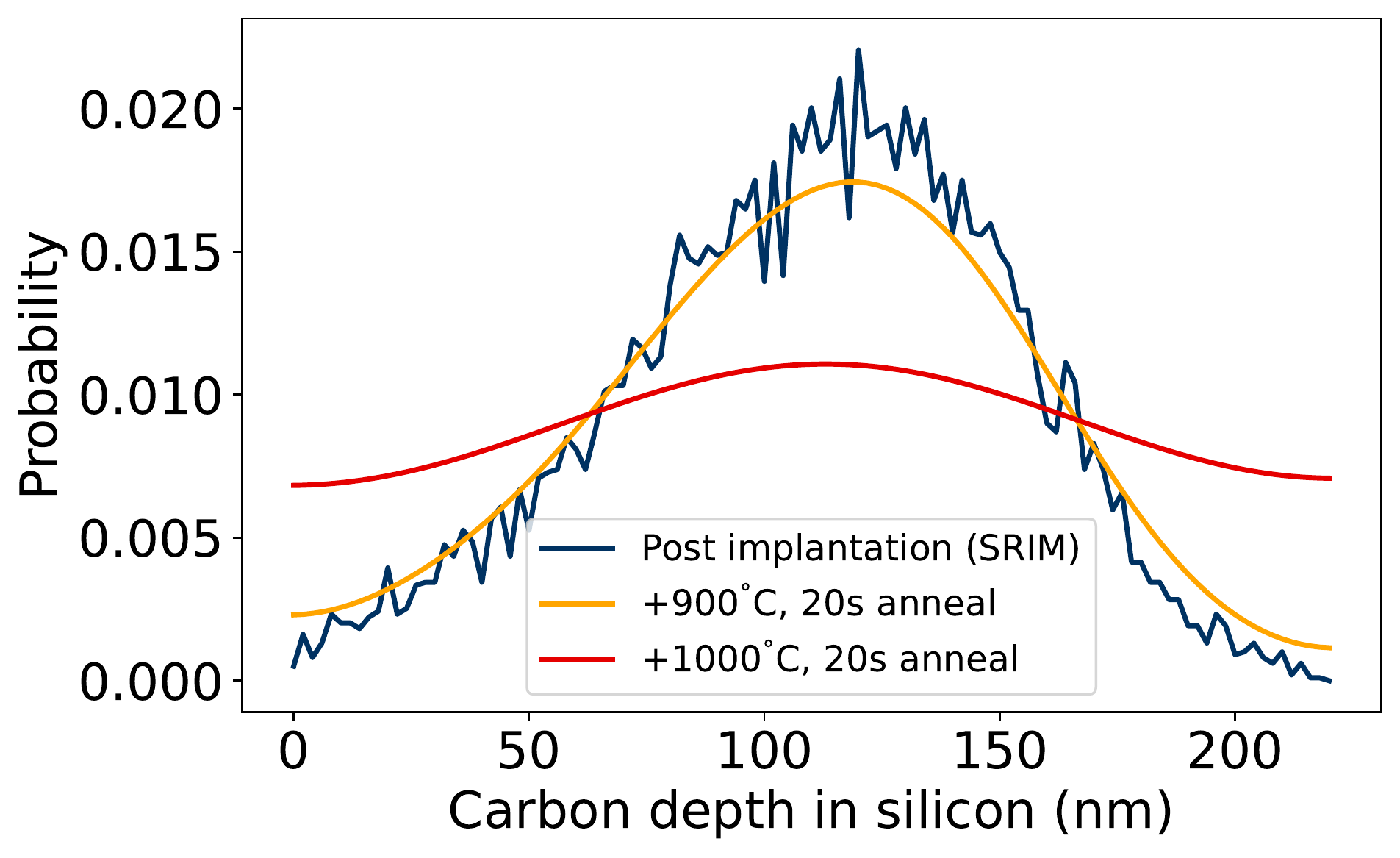}
    \caption{ Simulation of carbon distribution in silicon after implantation and rapid thermal annealing. In our experiments, we used a \SI{1000}{\degreeCelsius} thermal anneal for 20 s.
    Simulations based on diffusion coefficients in the literature suggest a near-uniform depth distribution of carbon inside the silicon waveguide.}
    \label{fig:diffusion}
\end{figure}

We used Stopping and Range of Ions in Matter (SRIM) simulations to estimate the depth distribution of carbon atoms in the silicon device layer after ion implantation at 36 keV and $7^\circ$ tilt. The SRIM simulations gave a mean depth of 112~nm and a longitudinal straggle of 41~nm. We used finite difference method to estimate the carbon distribution after annealing. We used a diffusion coefficient of carbon in silicon $D = 0.33~ e^{-2.92eV/kT}\textrm{cm}^2/\textrm{s}$ from Ref. \cite{Newman1961}. Our simulation results in Fig.~\ref{fig:diffusion} show that after our \SI{1000}{\degreeCelsius} thermal anneal for 20 s, the carbons are uniformly distributed inside the 220~nm device layer. These results indicate that choosing an annealing temperature of \SI{900}{\degreeCelsius} should maintain emitter localization near the center of the waveguide where the mode intensity is maximum. Further SIMS measurements are needed to develop an improved understanding of carbon diffusion and G center formation. These simulations suggest that the G center in our experiment could be positioned at any depth inside the waveguide. 

\section{First-principles modeling of G centers}
\label{firstprinciples}

\begin{figure}
    \centering
    \includegraphics[width=\columnwidth]{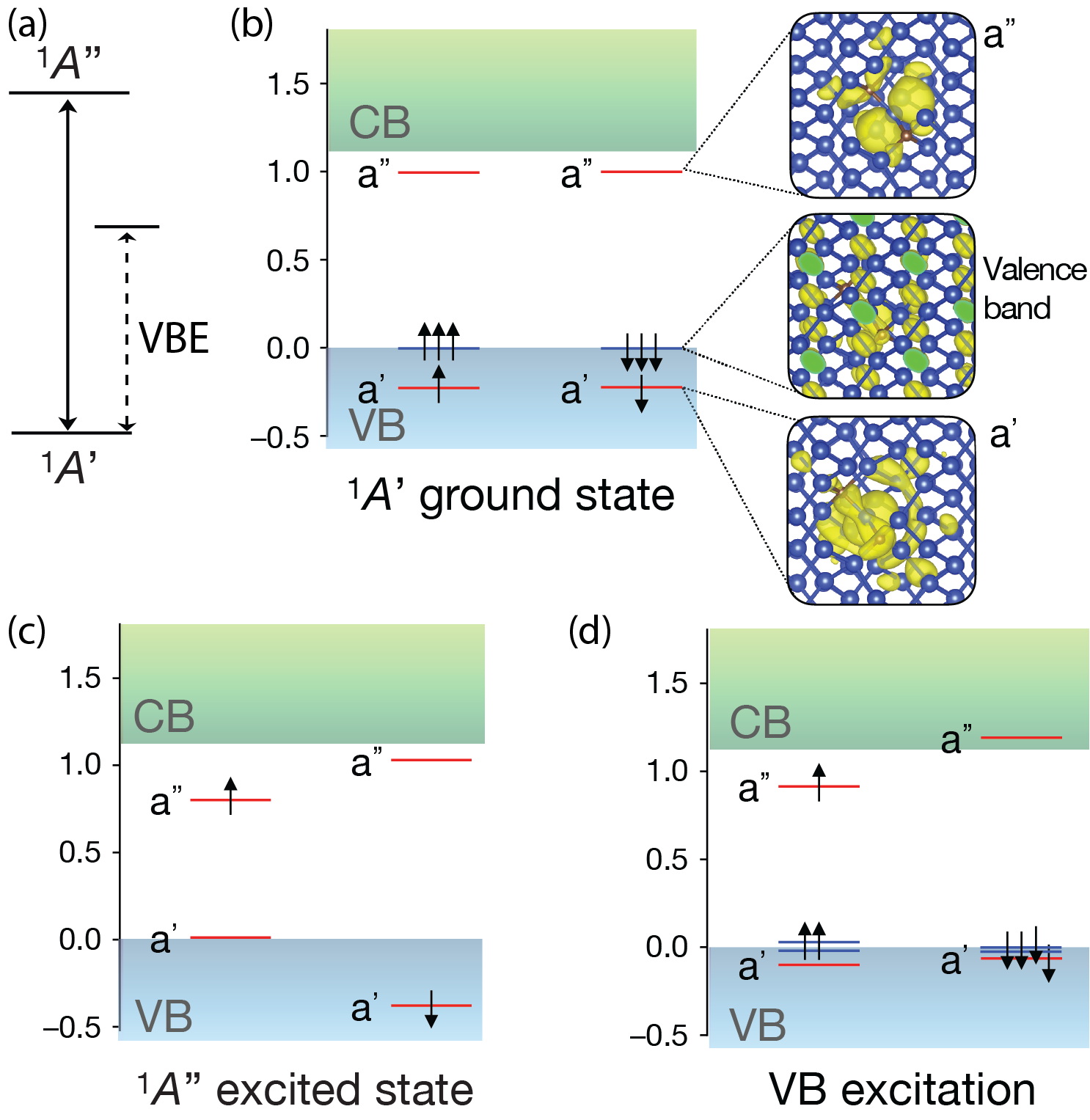}
 	\caption{Optical transitions in G centers from first principles. 
 	(a) Radiative transitions of the G center from the 1$A'$ ground state. 1$A'\leftrightarrow$ 1$A''$ are between localized defect levels. 
 	(b) The single-particle levels of the G center in its 1$A'$ ground state. The localized defect states ($a'$, $a''$) are shown in red, and the host valence band (VB) states are shown in blue.  	The localized orbitals $a'$ and $a''$ are shown as insets with an isocontour value of 0.0005 $a^{–3}$. The host valence band orbitals show delocalized character with an isocontour value of 0.0001 $a^{–3}$. 
 	(c) The 1$A''$ excited state is obtained by constraining the occupation of the unoccupied orbitals that creates the $a'$ hole state and populates the $a''$ state in the gap. (d) Radiative transition from a valence band state. The single-particle levels of G center where the radiative recombination is through the localized in gap-defect state $a''$ and the delocalized host state (blue).}
    \label{figS3:dft}
\end{figure}

We performed first-principles calculations using Vienna \textit{ab-initio} simulation package (VASP)~ \cite{G.Kresse-CMS96, G.Kresse-PRB96} with the projector augmented-wave method (PAW)~\cite{Blochl1994}. 
All calculations were spin-polarized with a plane wave cutoff energy of 400~eV. The Heyd-Scuseria-Ernzerhof (HSE) \cite{Heyd2003} functional with 25\% exact exchange was used to provide an improved description of the electronic structures to the semilocal functionals. The G center was positioned in a 512-atom supercell and a $\Gamma$-only k-point sampling. 
The supercell was optimized at a fixed volume until the forces on the ions were smaller than 0.01 eV/\AA. The single-particle Kohn-Sham levels of the G center at its 1$A'$ ground state are shown in Fig.~ \ref{figS3:dft}.

We used the so-called configuration B of the G center as multiple reports indicate that it is the configuration in best agreement with experimental data \cite{Ivanov2022, Gali2021, Song1990.PRB}. We found that in its ground state, the defect with symmetry C$\rm_{1h}$ introduces two highly localized defect levels where the $a''$ lying below the valence band maximum and the $a'$ within the band gap, 1 eV above the valence band. This single-particle picture agrees with other hybrid computations when taking into account slight differences in methodology and supercell size \cite{Ivanov2022, Gali2021, Wang2014.JAP}. It disagrees quantitatively with G$_0$W$_0$ which places the unoccupied state significantly lower in energy \cite{Timerkaeva2018.JAP}. We tentatively attribute this disagreement to the sensitivity of G$_0$W$_0$ to its starting wavefunctions obtained in the generalized gradient approximation (GGA).

We use constrained-HSE to simulate the 1$A''$ excited state of the G center as shown in Fig.~\ref{figS3:dft}. We have performed the excitation by emptying the localized $a'$ defect state below the valence band and occupying the $a''$ state. We noted that the resulting single-particle hole state $a''$ in the excited state moved slightly above the valence band edge, similar to negatively charged splitting vacancy in the diamond \cite{Gali2013,Gali-PRL2008}. The ZPL can be obtained from the energy difference between the total energy of the excited and the ground states. This methodology has been shown to give ZPLs within 100 meV from experiment for defects in diamond \cite{S.Li-naturecomm2022,Gali-PRL2008}. Our computed ZPL energy of the intra-defect transition is 1000 meV, which is in reasonable agreement with the experimental measurement of 968 meV in Fig.~\ref{fig2}. 
The ground state single-particle diagram (Fig.~\ref{figS3:dft}) suggests that an alternative excitation mechanism to an intra-defect transition would be to excite an electron from the valence band to the localized $a''$ defect state forming a bound exciton defect. Bound exciton defects have been suggested in the T center in silicon \cite{Bergeron2020,Dhaliah2022}. The computed ZPL for these valence band excitations are from 937 to 958 meV. These excitations show a much lower calculated transition dipole moment (0.65 to 0.83 D) than the intra-defect transition (2.4 to 3.3 D) due to the very different nature of the states (delocalized to localized). Accordingly, the computed radiative lifetime of the intra-defect transition is 0.15 to 0.3 $\mu$s, which is an order of magnitude smaller than that of the valence band excitations (3.59 to 3.69 $\mu$s). The intra-defect transition results in a much smaller radiative time in better agreement with the clear and bright PL of the G center. Our analysis suggests therefore that localized defect states from $a'$ to $a''$ are responsible for the PL of the G center. We note that emission from defect bound excitonic-like recombination to valence band states could be present in the PL spectra but not observable due to the phonon sideband and their much weaker signal. Additionally, our computations indicate that the transition dipole moment is aligned along the $\langle 110 \rangle$ direction.


\section{Efficiency analysis}

We define the system efficiency $\eta =0.4(1)\times10^{-3}$ as the probability of detecting a ZPL photon per excitation:
\begin{equation}
    \eta = \eta_{\text{\tiny QE}} \eta_\mathrm{wg} \eta_\mathrm{filter}^{\mathrm{BP}} \eta_\mathrm{network}
\end{equation}
where $\eta_{\text{\tiny QE}}$ is the quantum efficiency, $\eta_\mathrm{wg}$ is the probability of an excited G center emitting a photon into the waveguide mode, $\eta_\mathrm{filter}^{\mathrm{BP}}$ is the efficiency of our spectral filtering setup, and  $\eta_\mathrm{network}$ is the efficiency of our fiber network and detectors. We did not use the Fabry-Perot cavity during calibration measurements to reduce calibration uncertainties originating from the emitter linewidth. 
We assume that the emitter is excited with unit efficiency with each laser pulse.

We estimate the quantum efficiency using 

\begin{equation}
    \eta_{\text{\tiny QE}}=\frac{\eta} { \eta_\mathrm{wg} \eta_\mathrm{filter}^{\mathrm{BP}} \eta_\mathrm{network}}
\end{equation}
We simulate $\eta_\mathrm{wg}\leq0.8$, where the inequality is due to the uncertainty of the position of the emitter position. The filtering efficiency $\eta_\mathrm{filtering}\approx0.14$  is a product of the ZPL branching ratio (0.18) and bandpass filter transmission (0.8). We measure the efficiency of the remaining components in our setup to be $\eta_\mathrm{network}\approx0.2$, originating from a combination of losses in lensed fiber coupling efficiency (0.5), finite detector efficiency (0.6), and the remaining optical components in the fiber network.

From these calibrations, we can bound the quantum efficiency of the G center as $\eta_\text{\tiny QE}>0.02$, where the inequality is due to the uncertainty in the emitter position. We can put a corresponding upper bound on the radiative lifetime ($\tau_r=\gamma_r^{-1}<260$~ns) using: 
$$\eta_{\text{\tiny QE}} = \frac{\gamma_r}{\gamma_{nr} + \gamma_r} $$
where $\gamma_r$ and $\gamma_{nr}$ are the radiative and nonradiative decay rates, $\Gamma_G=\gamma_r+\gamma_{nr}=(4.6\textrm{ns})^{-1}$. We note that previous experiments \cite{Redjem2020telecom} found a tighter upper bound of $\tau_r<74$~ns. The estimated lifetime from our first principles calculations is between 150 and 300~ns.

\section{Time-resolved two-photon interference}
Our experimental two-photon interference results in Fig.~\ref{fig4} show that the successive photons emitted from a single G center in a photonic waveguide show a high degree of indistinguishability. We describe a model that explains the observed temporal dynamics which indicate that spectral fluctuations in G centers shows long-time correlations. 

We follow the time-resolved description of two-photon quantum interference in Ref.~\cite{Legero2003}, and adapt it for exponential wavepackets. We consider the time-resolved dynamics of two single-photon pulses simultaneously arriving at the beamsplitter (at $z=0$). These two pulses are created with a 25~ns delay at the source G center, and simultaneously arrive at the beamsplitter due to the delayed interforemeter setup shown in Fig.~\ref{fig4}. The two spatio-temporal mode function amplitudes for the photons at the position of the beamsplitter ($z=0$) are: 

\begin{align}
    \psi_{1,2} (t,z=0) &= \frac{1}{\sqrt{T_1}}\textrm{exp}(-t/(2 T_1)-i\omega_{1,2}t) 
    \label{eq:wavefunction}
\end{align}
where $\omega_1$ and $\omega_2$ are the carrier frequencies of the first and second single-photon pulses respectively, and $T_1$ is the emission lifetime as measured in Fig.~\ref{fig2}. The emission lifetime $T_1$ is a result of homogeneous broadening based on radiative $\gamma_r$ and non-radiative decay $\gamma_{nr}$ and is therefore constant between successive pulses. We consider the general case where the two photons can be at different frequencies. The origin of frequency differences between successive pulses is discussed below. 

The joint photon-detection probability due to the interference of two single-photon wavepackets is given by:

\begin{equation}
    P_\textrm{joint}(t_0,t_0+\tau)=\frac{1}{4}|\psi_1(t_0+\tau)\psi_2(t_0)-\psi_2(t_0+\tau)\psi_1(t_0)|^2
\end{equation}

For the exponential single-photon wavepackets in our experiment (Eq.~\ref{eq:wavefunction}, Fig.~\ref{fig2}), the joint photon detection probability becomes 
\begin{equation}
    P_\textrm{joint}(t_0,t_0+\tau)=\frac{e^{(-2t_0-\tau)/T_1}}{2T_1^2}(1-\textrm{cos}{\Delta\tau})
\end{equation}
and 
\begin{equation}
    P_\textrm{joint}(\tau)=\frac{e^{-\tau/T_1}}{4T_1}(1-\textrm{cos}{\Delta\tau})
    \label{eq:jointprob}
\end{equation}
upon integration. 
Photons can be made distinguishable at very large detunings ($\Delta\rightarrow\infty$) or by having orthogonal polarizations. In Eq.~\ref{eq:jointprob}, the distinguishable case corresponds to setting the cosine term equal to zero. We therefore define a parameter $\chi$ to describe the degree of indistinguishability of the single photons generated.

\begin{equation}
    P_\textrm{joint}(\tau)=\frac{e^{-\tau/T_1}}{4T_1}(1-\chi\textrm{cos}{\Delta\tau})=\alpha g^{(2)}(\tau)
    \label{eq:jointprobtau}
\end{equation}

where $\chi_\textrm{ideal}=1$ for indistinguishable and $\chi_\textrm{ideal}=0$ for distinguishable photons. We experimentally tune the indistinguishability by adjusting the relative polarization of the incoming photons to be parallel or orthogonal. This allows us to use experimentally measured values to quantify the degree of indistinguishability: 

\begin{equation}
    \chi_\textrm{exp}= 1-g^{(2)}_\parallel(0)/g^{(2)}_\perp(0) 
    \label{eq:visibility}
\end{equation}

Based on the description above, any detuning between the two photon pulses should still lead to high indistinguishability at $\tau=0$. However, large detunings result in very rapid oscillations that cannot be measured due to detector and electronics timing jitters. More importantly, it narrows the time interval and reduces the probability for successful coincidence detection. 

In the curve in Fig.~\ref{fig4}, we see an exponential feature around $g^{(2)}_\parallel(0)$ instead of a cosine as suggested by Eq.~\ref{eq:jointprobtau}. The experimentally observed exponential behavior near $|\tau|<1$ ns is caused by fluctuations in the detuning ($\Delta$) between successive pulses ($\psi_1,\psi_2$). For solid-state quantum emitters, the dominant source of spectral broadening ($\Gamma=\gamma_r/2+\gamma_{nr}/2+\gamma_d$) beyond the lifetime limit ($\gamma_r+\gamma_{nr}$) is caused by spectral diffusion. 
Spectral diffusion is a pure dephasing process ($\gamma_d$) where emission frequency fluctuates due to external classical noise sources such as fluctuating charges in the solid-state environment. Such charge fluctuations cause a frequency shift on the optical transition frequencies via the DC Stark shift \cite{Acosta2012}. Fig.~\ref{fig3}(a) shows that the  broadening in our system is well-captured by a model where the single-photon emission frequency is sampled from a Lorentzian distribution 

\begin{equation}
    p(\omega_i) = \frac{\Gamma}{2\pi}\frac{1}{(\omega_i-\omega_0)^2+(\Gamma/2)^2}
    \label{eq:linewidth}
\end{equation}

with a full width half maximum of $\Gamma/2\pi = 2.8$~GHz. If the emission frequency of two successive pulses are uncorrelated, the relative detuning $\Delta$ between successive pulses will be sampled from a Lorentzian distribution at twice the single-photon linewidth $\Gamma^{\textrm{uncorr}}=2\,\Gamma$. If the correlation timescale ($\tau_c$) of the emitted photon frequencies is longer than the two-photon delay ($\delta \tau=25$ ns in the experiment), we expect

\begin{equation}
    p(\Delta_{12}) = \frac{\Gamma^{\textrm{HOM}}}{\pi}\frac{1}{\Delta_{12}^2+(\Gamma^{\textrm{HOM}})^2} 
    \label{eq:linewidthHOM}
\end{equation}
where the effective two-photon linewidth in the experiment is $\Gamma^{\textrm{HOM}}<2\,\Gamma$. We assume $\Gamma\approx\gamma_d$ for simplicity since in the current experiments $\gamma_d$ is much greater than $\gamma_r+\gamma_{nr}$. 

Finally, we can obtain the experimentally observed $g^{(2)}(\tau)$ temporal dynamics by integrating Eq.~\ref{eq:jointprobtau} with the probability distribution for two photon detunings in Eq.~\ref{eq:linewidthHOM}

\begin{align}
    G^{(2)}(\tau) &= \int_{-\infty}^{\infty}\,d\Delta_{12}P_\textrm{joint}(\Delta_{12},\tau)p(\Delta_{12})\\
    &= \frac{e^{-\tau/T_1}}{4T_1} \left(1-\chi e^{-(\Gamma^{\textrm{HOM}}\tau)}\right) 
    \label{eq:modelfinal}
\end{align}

We fit the data to the functional form of $G^{(2)}(\tau) + \textrm{b.g.}$ where $\textrm{b.g.}$ is a variable to account for background noise.  We use the value of $T_1=\SI{4.6\pm0.1}{ns}$ from the lifetime measurements in Fig.~\ref{fig3}. We find $(\Gamma^\textrm{HOM})^{-1}= $\SI{0.4\pm0.1}{ns}. 

The effective two-photon linewidth in the quantum interference measurement is $\Gamma^\textrm{HOM}/2\pi= $\SI{0.4\pm1}{GHz} is about an order of magnitude smaller than the measured linewidth of the emission spectrum in Fig.~\ref{fig3}. We attribute this to the time dynamics of spectral diffusion in the system, as Fig.~\ref{fig3}(b) was acquired over a ten-minute period, while we interfere two subsequently emitted photons separated by 25~ns in our HOM experiment.

One significant factor limiting the depth of our HOM dip is the timing jitter of our SNSPDs. We measured this timing jitter by measuring laser-laser correlations and extracting $\sigma$ from a gaussian fit. We obtained a jitter of 252~ps. We include this in our fit model by convolving Eq.~\ref{eq:modelfinal} with a gaussian curve with $\sigma$ = 252~ps. In the main text, we report $g^{(2)}(0)$ and HOM visibility values based on the raw data, shot noise estimates, and Eq.~\ref{eq:visibility}, without any dependence on fit model details or timing jitter correction. 

\bibliography{alp,po,yihuang}

\end{document}